\documentstyle[emulateapj]{article}

\def\simlt{\lower.5ex\hbox{$\; \buildrel < \over \sim \;$}}
\def\simgt{\lower.5ex\hbox{$\; \buildrel > \over \sim \;$}}
\def\gsim{\;\rlap{\lower 2.5pt
 \hbox{$\sim$}}\raise 1.5pt\hbox{$>$}\;}
\def\lsim{\;\rlap{\lower 2.5pt
   \hbox{$\sim$}}\raise 1.5pt\hbox{$<$}\;}

\def\spose#1{\hbox to 0pt{#1\hss}}
\def\lta{\mathrel{\spose{\lower 3pt\hbox{$\mathchar''218$}}
     \raise 2.0pt\hbox{$\mathchar''13C$}}}
\def\gta{\mathrel{\spose{\lower 3pt\hbox{$\mathchar''218$}}
     \raise 2.0pt\hbox{$\mathchar''13E$}}}
\newcommand{\beq}{\begin{equation}}
\newcommand{\eeq}{\end{equation}}
\makeatletter
\newenvironment{tablehere}
  {\def\@captype{table}}
  {}
\makeatother

\lefthead{Haiman, Spaans \& Quataert}
\righthead{Cooling Radiation from High--Redshift Halos}

\begin{document}

\title{Lyman $\alpha$ Cooling Radiation from High--Redshift Halos}

\author{Zolt\'an Haiman\altaffilmark{1,4}, Marco Spaans\altaffilmark{2,4} \& Eliot Quataert\altaffilmark{3,5}}
\affil{$^{1}$Princeton University Observatory, Ivy Lane, Princeton, NJ 08544}
\affil{$^{2}$Harvard-Smithsonian Center for Astrophysics, 60 Garden Street MS 51, Cambridge, MA 02138}
\affil{$^{3}$Institute for Advanced Study, School of Natural Sciences, Olden Lane, Princeton, NJ 08540}
\altaffiltext{4}{Hubble Fellow}
\altaffiltext{5}{Chandra Fellow}

\authoremail{zoltan@astro.princeton.edu}
\authoremail{mspaans@cfa.harvard.edu}
\authoremail{eliot@ias.edu}

\vspace{\baselineskip}
\submitted{Submitted to ApJ Letters}

\begin{abstract}
The baryons inside high--redshift halos with virial temperatures above
$T\sim10^4$K cool radiatively as they condense inside dark matter
potential wells.  We show that the release of the gravitational
binding energy, over the halo assembly time--scale, results in a
significant and detectable Ly$\alpha$ flux.  At the limiting line flux
$\approx 10^{-19}~{\rm erg~s^{-1}~cm^{-2}~asec^{-2}}$ of the {\it Next
Generation Space Telescope}, several sufficiently massive halos, with
velocity dispersions $\sigma\gsim 120~{\rm km~s^{-1}}$, would be
visible per $4^\prime\times4^\prime$ field.  The halos would have
characteristic angular sizes of $\approx 10$\H{}, would be detectable
in a broad--band survey out to $z\approx 6-8$, and would provide a
direct probe of galaxies in the process of forming. They may be
accompanied by He$^+$ Ly$\alpha$ emission at the $\approx 10\%$ level,
but remain undetectable at other wavelengths. Our predictions are in
good agreement with the recent finding of two Ly$\alpha$ ``blobs'' at
$z=3.1$ by Steidel et al. (1999).
\end{abstract}

{\it subject headings}: cosmology:theory -- early universe -- galaxies: evolution -- galaxies: ISM

\section{Introduction}

In standard Cold Dark Matter (CDM) models for the formation of
structure in the universe, a significant population of DM halos
collapse at high redshift ($z\gsim 5$).  Provided that the gas inside
such a halo is heated to a virial temperature of $T\gsim 10^4$K (or
$T\gsim 10^2$K if ${\rm H_2}$ molecules are present, see, e.g.,
Haiman, Abel \& Rees 2000), it can cool on a short time--scale, and
continue condensing.  This condensation process can be regarded as a
cosmological cooling flow: a high--redshift and low--mass analogue of
cooling flows found in clusters of galaxies (e.g. Fabian 1994).

The behavior of the gas in early DM condensations should resemble the
formation of normal disk galaxies, which has been investigated by
several authors in semi--analytic schemes (White \& Rees 1978; Fall
\& Efstathiou 1980; Mo, Mao \& White 1998 [MMW]; van den Bosch \& 
Dalcanton 2000), as well as numerical simulations (Katz 1991; Navarro
\& White 1994; Navarro, Frenk \& White 1997 [NFW]; Navarro \&
Steinmetz 1997; Moore et al. 1999).  In all of these studies, most of
the baryons in the halo cool and settle into a rotationally supported
disk with an approximately exponential surface density profile.  In
the simplest picture, the disk material originates as smooth gas,
collapsing isothermally from the virial radius $R_{\rm vir}$ to its
final orbital radius of $\sim \lambda R_{\rm vir}$, where $\lambda\sim
0.05$ is the typical spin parameter. The numerical simulations reveal
a more complex process, where the gas forms smaller clumps early on;
these clumps then progressively merge together, collide, and
dissipate, to form the larger systems.

Irrespective of the dynamical details of the collapse, the disk
material must dissipate a minimum energy $\sim M_{\rm disk} v_c^2$,
where $M_{\rm disk}$ is the final mass of the disk and $v_c$ is the
circular velocity of the DM halo in which it is embedded.  The most
natural mechanism for this dissipation is radiation from collisionally
excited H and He, as well as Bremsstrahlung radiation for more massive
($T_{\rm vir}\gsim 10^6$K) DM halos.  Although this cooling radiation
has been assumed to be present in studies of the formation of
disk--galaxies, its observability has typically not been addressed.

In this {\it Letter}, we use a simple toy--model to quantify the
Ly$\alpha$ fluxes expected from the baryonic material cooling and
settling into the DM potential wells at high redshift.  We estimate
the number of sources that would be detectable with the {\it Next
Generation Space Telescope (NGST)}, and argue that these sources could
be identified based on their their large angular extent, relatively
shallow surface brightness profiles, and their low luminosity at other
wavelengths.  A detection of the cooling radiation would be a novel
and direct way to study proto--galaxies in the process of their
assembly.  Throughout this {\it Letter}, we assume the background
cosmology to be $\Lambda$CDM, with the parameters $\Omega_m=0.3,
\Omega_b=0.04, \Omega_\Lambda=0.7, h=0.65$, and $\sigma_8=1$.

\section{Cooling Radiation}

We first estimate the energy budget and total flux of the cooling
radiation from a halo under the following simplifying assumptions: (1)
initially, the baryons and the DM both follow the same density
profile, apart from a normalization by $\Omega_b/\Omega_m$, (2) the
baryons settle into a rotationally supported disk, and (3) the energy
radiated emerges as Ly$\alpha$ radiation.

A parcel of gas in a massive halo with virial temperature
significantly above $10^4$K would be shock--heated to a high
temperature, and initially cool via either He$^+$ line--cooling, or
via Bremsstrahlung.  This initial, fast cooling would release the
thermal energy of the gas, and reduce its temperature to $\sim
10^4$K.\footnote{This could result in additional He Ly $\alpha$ or
Bremsstrahlung cooling signals comparable to the H Ly$\alpha$
radiation discussed here.}  Since the $10^4$K gas is no longer
pressure supported, it contracts on a dynamical timescale, and moves
inward in the DM potential well. During this contraction, it is likely
to remain roughly isothermal, and radiate mostly Ly$\alpha$ radiation
(see discussion below).

The energy dissipated in this contracting phase must be approximately
the gravitational binding energy.  We now compute this energy
explicitly for the gas moving in the NFW potential.  We assume that
the gas and DM are initially both distributed according to
\begin{equation}
\rho(r) = \rho_0 \frac{1}{(r/r_s)(1+r/r_s)^{2}},
\label{eq:NFW}
\end{equation}
and that the final state is a rotationally supported exponential disk
embedded in the DM halo.  We use the solution for the disk surface
density $\Sigma(r)$ in MMW (their eq. 28) in the final configuration,
which takes into account the gravitational effect of the disk on the
halo in a self-consistent fashion.  A gas particle initially located
at radius $R_i$ settles into a circular orbit at a final radius $R_f$,
such that the initial and final enclosed gas masses are equal,
\begin{equation}
\int_0^{R_i} dr 4\pi r^2 \rho_{\rm gas}(r) = \int_0^{R_f} dr 2\pi r \Sigma_{\rm gas}(r).
\label{eq:radii}
\end{equation}
In the solution for $\Sigma_{\rm gas}(r)$, we adopt the typical
concentration index $c\equiv R_{\rm vir}/r_s=10$, spin parameter
$\lambda=0.05$, and assume that all baryons end up in the disk.  We
find contraction factors of $R_i/R_f\approx 10$, varying from a value
of $\approx 8$ for particles near the the surface $R_{\rm vir}$ to a
peak of $\approx 13$ near half the virial radius.  Note that more
realistically, a distribution of values for $\lambda$ should be used,
causing an intrinsic scatter in the luminosity of a halo of given
initial size.  This could enhance the number of halos visible at a
given Ly$\alpha$ flux.

The total energy released in cooling by a particle of mass $m$ in this
process is
\begin{equation}
\label{eq:Ecool1}
E_{\rm cool}= W-E_{\rm kin} = \int_{R_f}^{R_i} dr \frac{GM(r)m}{r^2}
 - \frac{GM(R_f)m}{R_f},
\end{equation}
where $W$ is the work done on the particle by the gravitational field,
$E_{\rm kin}$ is the kinetic energy of the particle in its final
circular orbit, and $M(r)$ is the total mass interior to
radius $r$.  Summed over all particles, we have
\begin{eqnarray}
\label{eq:work}
W= && \int_0^{R_{\rm vir}} dR_i 4\pi R_i^2 \rho_{\rm gas}(R_i) 
\int_{R_f}^{R_i} dr \frac{GM(r)}{r^2}\\
E_{\rm kin} = &&
\label{eq:kin}
\frac{1}{2}\int_0^\infty dr 2\pi r \Sigma(r) \frac{GM^\prime(r)}{r}.
\end{eqnarray}
Note that in eq.~\ref{eq:work}, we have used the initial gas
distribution $\rho_{\rm gas}$ and total (gas + DM) enclosed mass
$M(r)$. For simplicity, we have assumed that the initial NFW
distribution determines both quantities.  By contrast, in
eq. \ref{eq:kin}, $\Sigma(r)$ and $M^\prime(r)$ refer to the surface
density of the gaseous disk, and the total enclosed mass (DM+disk) in
the final equilibrium state, taken from the numerical solution in MMW.
Under these assumptions, we find that the total radiated energy is
\begin{equation}
\label{eq:rad}
E_{\rm cool}=W-E_{\rm kin}=1.9 E_{\rm bind},
\end{equation}
where $E_{\rm bind}=\int_0^{R_{\rm vir}}dr 4\pi r^2 \rho_{\rm gas}
GM(r)/r$ is the total initial gravitational binding energy.

In an early three--dimensional smooth particle hydrodynamics (SPH)
simulation, Katz (1991) explicitly computed the energy dissipated
during the collapse of a $7\times10^{11}~{\rm M_\odot}$ perturbation
in a standard CDM.  Adopting the parameters of this halo, we find the
energy radiated in our spherical approach to be $1.1\times 10^{59}$
erg.  This is only $\approx 30\%$ smaller than the value $1.5\times
10^{59}$ obtained in the simulation, lending support for our simple
estimates for the overall energy budget.

The timescale over which the dissipated energy is released is related
to the assembly of the halo.  In the simple picture of monolithic
collapse in a pre--existing spherical DM halo, the initial thermal
energy would be released quickly on a cooling time; subsequently, the
gas would contract isothermally on a dynamical time. Since the typical
overdensities are of order $\sim 100$; this would correspond to $\sim
1/\sqrt{100}=10\%$ of the Hubble time $t_{\rm Hub}(z)$.  However, in
the hierarchical structure formation scenarios of CDM cosmologies,
halos are built up gradually, as a result of collapse on smaller
scales, mergers, and continuous accretion of fresh gas.  The typical
time--scale on which this process occurs can be estimated within the
extended Press--Schechter ``merger tree'' formalism.  This formalism
yields the probability distribution of ages $dp/dt(M,z,t_h)$ for halos
of mass $M$ existing at redshift $z$, where the characteristic
``age'', $t_h$, is defined to be the time elapsed since the halo had
acquired half of its present mass (Lacey \& Cole 1993, eq. 28).

For a halo of mass $M$ at redshift $z$, we adopt the definition
\begin{equation}
\label{eq:tdis}
\int_0^{t_d} dt \left[\frac{dp}{dt}\left(M,z,t\right)\right] =0.5
\end{equation}
for the characteristic time $t_d$ over which the gas assembles and
dissipates its energy. This definition ensures that half of the halos
build up half of their final mass within the time $t_d$.  The time
$t_d$ is a fraction of the Hubble time and depends on the halo
mass. On average, larger, rarer halos are assembled more recently than
typical halos corresponding to $1\sigma$ density perturbations.  In
general, for the halo masses relevant for our discussion, we find
$0.15\lsim t_d/t_{\rm Hub}(z) \lsim 0.30$, slightly longer than the
dynamical time.

\section{Detectability of Lyman $\alpha$ Halos}

Armed with the energy budget and characteristic dissipation time, we
can now make simple estimates for the expected Ly$\alpha$ fluxes from
gas cooling in halos.  The total line flux from a halo of mass $M$ at
redshift $z$ is simply given by
\begin{equation}
\label{eq:flux}
F_\alpha(M,z) = \frac{E_{\rm cool}(M,z)/t_d(M,z)}{4\pi d^2_l(z)},
\end{equation}
where $d_l(z)$ is the luminosity distance to redshift $z$.  A halo
will subtend an angle $2\theta_{\rm vir}=2R_{\rm vir}/d_{\rm A}\approx
12$\H{}$(R_{\rm vir}/50~{\rm kpc}) [d_{\rm A}(z=5)/d_{\rm A}(z)]$, and
will be easily resolved at {\it NGST}'s angular resolution of $\approx
0.05$\H{}.

To assess the observability of our sources, we use the proposed
instrumental characteristics of {\it NGST} with an 8m diameter
mirror. The expected signal--to--noise ratio in an observation of
duration $t$ is
\begin{equation}
\label{eq:sn}
\frac{S}{N}= \frac{A f t F_\alpha(1+z)/h\nu_\alpha}{
(A f t \Delta\lambda \pi \theta_{\rm vir}^2 I_{\rm bg})^{1/2}},
\end{equation}
which yields approximately
\begin{eqnarray}
\label{eq:sn2}
\nonumber
\frac{S}{N}\approx && 5
\left(\frac{A}{\rm 50~m^2}\right)^{1/2}
\left(\frac{f}{\rm 0.5}\right)^{1/2}
\left(\frac{t}{\rm 2.5\times10^5~s}\right)^{1/2}
\left(\frac{R}{\rm 5}\right)^{1/2}\\
&&
\times
\left(\frac{2\times10^4}{I_{\rm 1.25}}\right)^{1/2}
\left(\frac{M}{\rm 3\times10^{11}~{\rm M_\odot}}\right)^{4/3}
\left(\frac{1+z}{\rm 6}\right)^{2}
\end{eqnarray}
where $A$ is the mirror area, $f$ is a fiducial system throughput,
$R\equiv\Delta\lambda/\lambda_{\rm obs}$ is the assumed filter width,
$t$ is the integration time, and $I_{\rm bg}=I_{\rm
1.25}(\lambda/1.25\mu{\rm m})^{-1.5}~{\rm
photons~cm^{-2}~s^{-1}~sr^{-1}~\AA^{-1}}$ is the zodiacal background,
extrapolated to wavelengths shorter than $1.25~{\rm \mu m}$ from
Kelsall et al. (1998).  The background turns over at $\lambda\lsim
0.5\mu$m (Leinert et al. 1998), so that our extrapolation is an
overestimate at these wavelengths\footnote{See the {\it NGST} exposure
time calculator at http://augusta.stsci.edu for more details.}. We
have approximated the angular diameter distance as $d_l\propto
(1+z)^{0.75}$, and the dissipation time as $t_{\rm d}\propto
(1+z)^{-1.5}$.  These scalings are accurate to within a factor of two
in the relevant mass and redshift range; in our calculations we use
the exact expressions.

The expected line--widths are of order $\sim 30[(1+z)/6]\AA$ (see
below), and a narrow band filter with a comparable width would
minimize the sky background.  In an initial blind search for the
Ly$\alpha$ halos, however, we find that it is more advantageous to use
a broader filter, which covers a larger redshift range.
Eq.~\ref{eq:sn2} reveals that the minimum detectable halo mass scales
with the filter width approximately as $M\propto
\Delta\lambda^{3/8}$; the total number of halos visible per
observation then scales as $N\propto dz M(dN/dM)
\propto \Delta\lambda/M \propto \Delta\lambda^{5/8}$.  Accordingly,
in the estimates below we adopt a broad--band filter width of
$\Delta\lambda/\lambda_{\rm obs}=0.2~(R=5)$.  The limiting surface
brightness near the wavelength of $1\mu$m in a $2.5\times10^5$ second
observation, is $F_{\rm min}\approx 10^{-19}~{\rm
erg~s^{-1}~cm^{-2}~asec^{-2}}$ (S/N=5).  In comparison, the limiting
line fluxes in current searches for high--$z$ Ly$\alpha$--emitters are
typically $\approx 10^{-17}~{\rm erg~s^{-1}~cm^{-2}}$.

In Table~\ref{tab:list}, we list examples for the detectability of
Ly$\alpha$ halos at NGST's limiting line flux at different redshifts,
and summarize the properties of halos at the detection threshold.  In
order to produce the required flux, halos located at redshifts $2<z<8$
must have circular velocities of around $170~{\rm km~s^{-1}}$, or
total masses above $M_{\rm min}\approx 10^{11}~{\rm M_\odot}$.  Note
that because of the smaller angular sizes at higher redshift, and
because of the decreasing background at longer wavelengths, a {\it
smaller} halo is detectable at higher redshifts. Interestingly, the
circular velocity corresponding to the detection threshold is nearly
independent of redshift.  This is apparent from eq.~\ref{eq:sn2},
since $v_c^2\propto M^{4/3}(1+z)^2$. In order to be detectable at
$z\approx 3$ using current instruments, the required circular velocity
is larger, $\approx 300~{\rm km~s^{-1}}$, corresponding to a total
mass near $7\times 10^{12}~{\rm M_\odot}$. In the last column of
Table~\ref{tab:list}, we have shown the surface density of halos more
massive than $M_{\rm min}$,
\begin{equation}
\label{eq:dndz}
N_{\rm tot}= \Delta z \Delta\Omega
\frac{dV}{dzd\Omega}
\frac{t_d}{t_{\rm Hub}(z)}
\int_{M_{\rm min}(z)}^\infty dM \frac{dn}{dM} 
\label{eq:dNdzdom}
\end{equation}
where $dV/dzd\Omega$ is the cosmological volume element, $dn/dM$ is
the mass--function of halos taken from the Press--Schechter (1974)
formalism, $t_d/t_{\rm Hub}(z)$ is the duty--cycle,
$\Delta\Omega=4^\prime\times4^\prime$ is the proposed field of view of
{\it NGST}, and $\Delta z = 0.2(1+z)$ is the redshift range probed by
our filter.  As the table shows, $\sim 30$ halos are detectable within
each {\it NGST} survey field, with $\sim$1 halo still visible from a
redshift of $z\approx6-7$.  It should be kept in mind that much
fainter halos at higher redshifts could be detectable through a narrow
filter.  For instance, adopting a width $R=1,000$ would allow a
detection of a halo with $M\approx 10^{10}~{\rm M_\odot}$, $v_{\rm
c}=90~{\rm km~s^{-1}}$ out to $z=10$ (S/N=5, $t=2.5\times10^5$ s).

At the sensitivity threshold of current instruments, the predicted
surface density of Ly$\alpha$ halos near $z=5$ is $\approx 600$ per
square degree.  This value is $\approx 4\%$ of the surface density of
Ly$\alpha$ emitters at $z=4.5$ found in a recent narrow--band search
with the Keck Telescope (Hu et al. 1998).  The last line of
Table~\ref{tab:list} shows the number of halos expected at the
sensitivity level of the Steidel et al. (1999) data.  They found two
Ly$\alpha$ ``blobs'', compared to our prediction of $\sim 0.4$ such
objects in a 78 sq. amin field with a three--dimensional overdensity
of $15$, and in a redshift interval of $\Delta z=0.06$.  Their halos
have angular sizes of $15$\H{} and $17$\H{}, which are $\approx 50\%$
smaller than the size we obtained from the virial radius.

\begin{tablehere}
\vspace{0.2cm}
\caption{\label{tab:list} \footnotesize Examples for the detectability
and surface density of Ly$\alpha$ halos.}
\vspace{0.1cm}
\noindent\begin{tabular}{|c||c|c|c|c|c|c|c|}
\hline
$z$ & $\frac{M_{\rm min}}{\rm 10^{11} M_\odot}$ & $\frac{t_d}{t_{\rm
Hub}}$ & $\frac{T_{\rm vir}}{\rm 10^6 K}$ & $\frac{v_c}{\rm km/s}$ &
$\frac{F_\alpha/10^{-17}}{\rm erg/s/cm^2}$ & $\frac{\theta_{\rm
vir}}{\rm asec}$ & $\frac{N_{\rm tot}}{4^\prime\times4^\prime}$ \\
\hline
  2 & $7.5$ & 0.34 & 2.0 & 166 & $7.96$ & 15.4 & 11.2   \\
  3 & $5.2$ & 0.28 & 2.1 & 170 & $3.97$ & 11.0 & 9.1    \\
  4 & $3.9$ & 0.24 & 2.2 & 171 & $2.41$ & 8.8  & 5.7    \\
  5 & $3.0$ & 0.22 & 2.2 & 172 & $1.63$ & 7.4  & 3.1    \\
  6 & $2.3$ & 0.19 & 2.2 & 171 & $1.16$ & 6.5  & 1.5    \\
  7 & $1.9$ & 0.18 & 2.1 & 170 & $0.87$ & 5.7  & 0.7    \\
  8 & $1.6$ & 0.16 & 2.1 & 169 & $0.69$ & 5.2  & 0.3    \\
\hline
3.1 & $70$  & 0.22 & 12  & 400 & $3.7$  & 25   & 0.07   \\
\hline
\end{tabular}
\vspace{\baselineskip}
\end{tablehere}

\section{Further Considerations}

\subsection{Details of Collapse and Cooling}

The Ly$\alpha$ cooling signal proposed here is independent of the fate
of the dissipating gas in the final stages of collapse, i.e. the
details of disk--formation, fragmentation, and star--formation.
Nevertheless, a few caveats must be kept in mind.  Three--dimensional
simulations of galaxy formation have shown that a galactic halo is
built up by mergers of pre--existing subclumps, rather than by
continuous accretion of smooth gas.  At the current resolution of
simulations the fraction of the infalling mass in these clumps is
$\approx 10\%$ (e.g. Yoshida et al. 2000).  Since the slope of the
subclump mass--function is $dN/dM \propto M^{-1.8}$, the simulations
might already have converged on the total clumped mass.  Even stronger
sub--clumping, however, should not effect the overall energy budget:
as long as the gas contracts and maintains its circular velocity, it
has to dissipate the energy given in eq.~\ref{eq:rad}.  Furthermore,
the gas in small sub--clumps can be stripped from the merging DM
sub--clumps, either by the external UV background, or by clump--clump
collisions.  The baryons falling into the halo are then dispersed into
the inter--clump medium and shock--heated, implying that their
contraction and dissipation closely resembles the smooth accretion
scenario.  The strength of the internal shocks are determined by the
streaming velocities, which numerical simulations find to be
significantly smaller than the velocity dispersion of the halo. As a
consequence, the internal shocks are not likely to heat the gas to
temperatures above $T\approx 6\times 10^4$K, where He$^+$ line cooling
would start to dominate over Ly$\alpha$ cooling.

We have assumed that the dissipated energy is released in Ly$\alpha$
cooling alone. Indeed, for a pure H+He gas of primordial composition,
this is a safe assumption, since cooling shuts off abruptly below the
collisional excitation temperature of $T\approx 10^4$K.  However, in
the presence of molecules and heavy elements, radiative cooling is
possible below $10^4$K.  The cooling time for Ly$\alpha$ cooling at
the {\it peak} of the cooling curve is given by $t_{\rm cool}= (3/2) n
k_{\rm B} T_{\rm vir} / n^2 \Lambda_{\rm cool}$, where $\Lambda_{\rm
cool}\approx 2\times 10^{-22}~{\rm erg~s^{-1}~cm^{3}}$ and $n$ is the
baryon number density corresponding to an overdensity of $\approx 200$
at the time of collapse.  We have used $T_{\rm vir}$ in this
expression, rather than the gas temperature $T\approx 10^4$K, since it
is the former that characterizes the gravitational binding energy that
is radiated away.  We find that the cooling time is between $6-45\%$
of the dynamical time, $t_{\rm dyn}=1/\sqrt{6\pi G\rho}$ for the halos
considered here.  The requirement that $t_{\rm cool}\approx t_{\rm
dyn}$ then implies that the gas temperature will be near, but somewhat
below, the temperature $T\approx10^{4.1}$K where the cooling function
peaks.

Now consider the cooling of metal--enriched gas.  Although
non-Ly$\alpha$ cooling is possible at $T<10^4$K, the cooling function
below this temperature drops sharply. As long as the metalicity is
$Z<0.1 Z_\odot$, molecular cooling or cooling by atoms heavier than He
is at least a factor of $\approx 1000$ less efficient then cooling via
the Ly$\alpha$ peak. Since initially the non-Ly$\alpha$ cooling time
exceeds the dynamical time by a factor of $60-450$, and $t_{\rm
cool}/t_{\rm dyn}\propto \rho^{-1/2}$, Ly$\alpha$ cooling dominates
until the density is enhanced by a factor of $4\times10^3 -
2\times10^5$, or until radial contraction factors of $15-60$.  We
therefore conclude that the bulk of the binding energy is likely
released in Ly$\alpha$ cooling, until the metalicity builds up to
near-solar levels.  

Dust absorption can strongly suppress the Ly$\alpha$ flux escaping
from a medium that is optically thick to Ly$\alpha$ photons; this is
thought to be the reason why early Ly$\alpha$ surveys did not detect
proto--galaxies (see Pritchet 1994).  Recently, however, Ly$\alpha$
emitting galaxies have been found at high--redshift (e.g. Hu et
al. 1998), as expected in models with lower galactic dust abundance,
and inhomogeneous dust distribution (Haiman \& Spaans
1999). Furthermore, the dust abundance in the early, spatially
extended, collapsing phase of the high--redshift halos is likely to be
significantly lower than inside star--forming galaxies.

The typical Ly$\alpha$ luminosities we derive are $\sim
10^{43-44}~{\rm erg~s^{-1}}$. Our halos are expected to be mostly
neutral, and since they are embedded in the UV background, they will
convert a fraction $\approx 0.5$ of the incident ionizing radiation to
Ly$\alpha$ radiation (Gould \& Weinberg 1996).  It is interesting to
note that for a UV background flux of $5\times
10^{-22}~(h\nu/13.6~{\rm eV})^{-1.7}~{\rm
erg~s^{-1}~cm^{-2}~Hz^{-1}~sr^{-1}}$, the resulting Ly$\alpha$
radiation would be at a level of $5-10\%$ of our cooling signal.

\subsection{Contribution of Ly$\alpha$ Halos to the UV Background}

In addition to the Ly$\alpha$ flux expected from individual halos, we
have computed the contribution to the extragalactic UV background from
halos below the detection threshold, by summing the flux of these
sources.  We find the amplitude for this background to be quite small,
at the level of $\lsim 1\%$ of the UV background of 20$\pm10~{\rm
nW~m^{-2}~sr^{-1}}$ (Bernstein 1997).  Note, however, that the
Ly$\alpha$ background from collapsing halos should always dominate at
sufficiently high redshifts, when other UV sources, such as stars or
quasars, have not yet formed.

\subsection{Line Profiles: Shape and Interpretation}

In order to maximize the diagnostic value of observations of the
Ly$\alpha$ halos, it is important to understand the frequency
distribution and surface brightness profiles over which the Ly$\alpha$
photons are emitted.  In a naive spherical model, one may assume a
steady emission of Ly$\alpha$ photons over the halo assembly time. For
a purely radial contraction inside the logarithmic potential of an
isothermal sphere, the luminosity of each shell would scale
approximately as $L(r)\propto \int r^2 dr (nk_{\rm B}T_{\rm
vir})/t_d\propto \int r^2 dr n^{3/2}\propto \log r$. Most of the
luminosity would arise near the outer shells, and the surface
brightness profile would scale approximately as $I(r)\propto r^{-2}$
out to $\approx 30\%$ of the virial radius.  A similar conclusion
holds for the NFW profile, but with $I(r)$ flattening to $I(r)\propto
r^{-1}$ near $r=0$.

Although speculative, these surface brightness distributions have
interesting consequences.  Since they are peaked towards $r=0$, this
could potentially make the Ly$\alpha$ halos more observable than our
estimates in Table~\ref{tab:list} imply.  For example, a fraction
$\approx 1/3$ of the total luminosity is released internal to the
radius $R_{\rm vir}/10$.  Taking the total flux to be $1/3$ of the
value in eq.~\ref{eq:flux}, and assuming an extended source with an
angular size of $\theta_{\rm vir}/10$ significantly increases the
number of detectable sources.  At redshift $z=5$, the inner $\approx
0.6$\H{} region of a $v_c=130$ km/s halo would still be detectable,
providing $\approx 12$ candidate sources per {\it NGST}
field. Similarly, the innermost $\approx 0.3$\H{} region of a halo
with $v_c=120$ km/s would be visible to $z\approx 10$.

The depth of the DM potential well would imply a line--width of
$\approx 200~{\rm km~s^{-1}}$.  The line profile is, however,
significantly modified by scattering effects. Assuming that the halo
is mostly neutral, the Ly$\alpha$ optical depth at the center of the
line is $\tau= n_H \sigma_0 R_{\rm vir}\approx 10^7$.  For a thermal
speed of $\sim 10$ km s$^{-1}$ it requires a shift in frequency by an
amount $\delta\lambda/\lambda\approx 3\times 10^{-3}$ to diffuse out
of the optically thick core of a Voigt-profile (i.e. to a frequency
where $\tau=1$).  This will significantly broaden the line, to $\sim
1,000~{\rm km~s^{-1}}$; it would also further flatten the surface
brightness profile. Depending on the infall velocity and neutral
density structure, the line may exhibit a characteristic
``double--hump'' profile.

It would be very interesting to extract these quantities directly from
a three--dimensional simulation in the future, since the angular
resolution and spectral capabilities of {\it NGST} should be
sufficient to map the surface brightness distribution, and to
determine the line profile, at least for the brightest halos.  Here we
simply note that the characteristic line--widths of $\sim 1000~{\rm
km~s^{-1}}$, as well as the extended, and relatively shallow surface
brightness profiles, would be robust indicators of the initial stages
of the collapse.

\section{Discussion and Conclusions}

We have considered the Ly$\alpha$ flux from baryonic gas condensing
and cooling inside high--redshift galactic DM halos.  Since the
hydrogen in the universe is reionized beyond a redshift $z\approx 6$,
the cooling radiation would not suffer strong scattering in the
neutral IGM below this redshift (cf. Rybicki \& Loeb 1999), but the
halos would still appear extended on the sky with an angular diameters
of $\approx 10$\H{}.  We have ignored the radiation of the initial
thermal energy of the gas.  Most of this energy is radiated in the
continuum as Bremsstrahlung, but is too faint to be observable with
{\it CXO} or {\it XMM}.  However, for sufficiently metal--poor halos,
$\approx 10\%$ of the initial thermal energy could be released by
He$^+$ Ly$\alpha$ cooling.  This emission likely occurs on the
halo--assembly time--scale, and would accompany our hydrogen
Ly$\alpha$ signal at a wavelength 4 times shorter, and with a
$\approx10$ times lower line flux.

It is interesting to compare the overall binding energy radiated in
Ly$\alpha$ photons by the pristine halos and the stellar Ly$\alpha$
known to be produced in ``normal'' galaxies.  The total specific
energy radiated in Ly$\alpha$ is $E({\rm cool})\approx 2v_c^2$, where
$v_c$ is the circular velocity. According to Kennicut (1998), 1 solar
mass of gas cycled through stars releases $3.1\times 10^{49}$ erg in
Ly$\alpha$. Assuming that $10\%$ of the gas is converted into stars,
we find a specific energy of $E({\rm stars})=1.7\times 10^{-6}c^2$.
As a result, for a $v_c=200$km/s halo we find $E({\rm cool})/E({\rm
stars})= 0.5$. This relatively large ratio demonstrates that the total
flux in the cooling radiation can be comparable to that in stellar
Ly$\alpha$ emission lines.

Our predictions for the fluxes, angular sizes, characteristic
line--width, surface brightness profiles, and the number of sources
are in good agreement with the two Ly$\alpha$ ``blobs'' discovered at
$z=3.09$ in a 78 sq. amin field by Steidel et al. (1999), which leads
us to believe that examples of cooling Ly$\alpha$ halos may already
have been found.  Our simple spherical model, however, predicts a
``double--hump'' line profile, which is not seen by Steidel et al.
Further work is required to assess whether the 3D structure of the
collapsing halo can wash out the predicted double peak. 

Our results show that a few dozen cooling halos could be detected in
the future out to $z\approx 6-8$ in a single $4^\prime\times4^\prime$
field, observed for 2.5 ksec with {\it NGST}.  In order to extend the
redshift range over which this study is feasible, a large area survey
with {\it NGST} would be desirable, with a relatively broad filter (to
capture rarer, higher-$z$ halos, over a larger redshift range), as
well as wavelength coverage extending down to $\approx0.4\mu$m (to
capture halos at redshifts as low as $z\approx 3$).  Using a
narrow--band filter that minimizes the background would allow the
detection of additional, fainter sources, and a detailed study of the
brighter halos.  The cooling halos would be a novel and direct probe
of galaxy formation in the high--redshift universe, and of their
evolutionary history from $z\approx 6-8$ to lower redshifts.

\acknowledgments 

ZH thanks the hospitality of the ITP at the University of California
at Santa Barbara, where some of this work was carried out. The authors
gratefully acknowledge discussions with T. Abel, L. Bildsten,
C. Steidel, M. Steinmetz, N. Katz, C. Kochanek, and R. Bernstein.
This research was supported by NASA through Hubble Fellowship grants
HF-01101.01-97A (to MS) and HF-01119.01-99A (to ZH), and through
Chandra Fellowship grant PF9-10008 (to EQ), and by the NSF under
Grant No. PHY94-07194 at the ITP.


\begin{thebibliography}{}
\bibitem[]{bernstein97} Bernstein, R. 1997, PhD thesis, Caltech
\bibitem[]{f94} Fabian, A. C. 1994, ARA\&A, 32, 277
\bibitem[]{fe80} Fall, M. S., \& Efstathiou, G. 1980, MNRAS, 193, 189
\bibitem[]{gw96} Gould, A., \& Weinberg, D. 1996, ApJ, 468, 462
\bibitem[]{har00} Haiman, Z., Abel, T., \& Rees, M. J. 2000, ApJ, in press, astro-ph/9903336
\bibitem[]{hs99} Haiman, Z., \& Spaans, M. 1999, ApJ, 518, 138
\bibitem[]{hu98} Hu, E.M., Cowie, L.L., \& McMahon, R.G.\ 1998, ApJL, 502, 99
\bibitem[]{k91} Katz, N. 1991, ApJ, 368, 325
\bibitem[]{kel98} Kelsall, T. et al. 1998, ApJ, 508, 44
\bibitem[]{ken98} Kennicutt, R.C., Jr.\ 1998, ApJ, 498, 541
\bibitem[]{lc93} Lacey, C., \& Cole, S. 1993, MNRAS, 262, 627
\bibitem[]{l98} Leinert, Ch. et al. 1998, A\&ASS, 127, 1
\bibitem[]{mo98} Mo, H.J., Mao, S. \& White, S.D.M.\ 1998, MNRAS, 295, 319 [MMW]
\bibitem[]{m99} Moore, B., Ghigna, S., Governato, F., Lake, G., Quinn, T., Stadel, J. \& Tozzi, P. 1999, ApJL, 524, 19
\bibitem[]{nfw97} Navarro, J. F., Frenk, C. S., \& White, S. D. M. 1997, ApJ, 490, 493 [NFW]
\bibitem[]{ns97} Navarro, J. F., \& Steinmetz, M. 1997, ApJ, 478, 13
\bibitem[]{nw94} Navarro, J. F., \& White, S. D. M. 1994, MNRAS, 267, 401
\bibitem[]{ps74} Press, W. H., \& Schechter, P. L. 1974, ApJ, 181, 425
\bibitem[]{p94} Pritchet, C. J. 1994, PASP, 106, 1052
\bibitem[]{rl99} Rybicki, G.B. \& Loeb, A.\ 1999, ApJ, 524, 527
\bibitem[]{shap99} Shapiro, P. R., Ilyev, I., \& Raga, A. C. 1999, MNRAS, 307, 203
\bibitem[]{st99} Steidel, C.C., Adelberger, K, Shapley, A. E., Pettini, M.,  Dickinson, M., \& Giavalisco, M.\ 1999, ApJ, in press, astro-ph/9910144
\bibitem[]{vd00} van den Bosch, F. C., \& Dalcanton, J. J. 2000, ApJ, in press, astro-ph/9912004
\bibitem[]{wr78} White, S. D. M., \& Rees, M. J. 1978, MNRAS, 183, 341
\bibitem[]{y00} Yoshida, N., Springel, V., White, S. D. M. , \& Tormen, G. 2000, ApJL, submitted, astro-ph/0002362
\end{thebibliography}
\end{document}